**Moderate Resolution *Spitzer* Infrared Spectrograph (IRS) Observations of M, L, and T Dwarfs**

A. K. Mainzer[1], Thomas L. Roellig[2], D. Saumon[3], Mark S. Marley[4], Michael C. Cushing[5, 6], G. C. Sloan[7], J. Davy Kirkpatrick[8], S. K. Leggett[9], John C. Wilson[10]

[1] Jet Propulsion Laboratory, MS 169-506, 4800 Oak Grove Drive, Pasadena, CA 91109, amainzer@jpl.nasa.gov
[2] NASA Ames Research Center, MS 245-6, Moffett Field, CA 94035, thomas.l.roellig@nasa.gov
[3] Los Alamos National Laboratory, Applied Physics Division, MS P365, Los Alamos, NM 87545, dsaumon@lanl.gov
[4] NASA Ames Research Center, MS 254-3, Moffett Field, CA 94035, mmarley@mail.arc.nasa.gov
[5] Steward Observatory, University of Arizona, 933 North Cherry Avenue, Tucson, AZ 85721, mcushing@as.arizona.edu [6]Spitzer Fellow
[7] Astronomy Department, Cornell University, Ithaca, NY 14853, sloan@isc.astro.cornell.edu
[8] Infrared Processing and Analysis Center, MC 100-22, California Institute of Technology, Pasadena, CA 91125, davy@ipac.caltech.edu
[9] Gemini Observatory, 670 N. A'ohoku Place, Hilo, HI 96720, sleggett@gemini.edu
[10] Astronomy Building, University of Virginia, 530 McCormick Road, Charlottesville, VA 22903, jcw6z@virginia.edu

**ABSTRACT**

We present 10 – 19 μm moderate resolution spectra of ten M dwarfs, one L dwarf, and two T dwarf systems obtained with the Infrared Spectrograph (IRS) onboard the *Spitzer Space Telescope*. The IRS allows us to examine molecular spectroscopic features/lines at moderate spectral resolution in a heretofore untapped wavelength regime. These $R=\lambda/\Delta\lambda \sim 600$ spectra allow for a more detailed examination of clouds, non-equilibrium chemistry, as well as the molecular features of $H_2O$, $NH_3$, and other trace molecular species that are the hallmarks of these objects. A cloud-free model best fits our mid-infrared spectrum of the T1 dwarf ε Indi Ba, and we find that the $NH_3$ feature in ε Indi Bb is best explained by a non-equilibrium abundance due to vertical transport in its atmosphere. We examined a set of objects (mostly M dwarfs) in multiple systems to look for evidence of emission features, which might indicate an atmospheric temperature inversion, as well as trace molecular species; however, we found no evidence of either.

## 1. Introduction

Since the discovery of the first brown dwarf (BD) Gl 229B (Nakajima et al. 1995) and the confirmation of other BD candidates (Becklin & Zuckerman 1988; Basri et al. 1996; Rebolo et al. 1996), extensive efforts have been made to characterize the atmospheric features of these objects both theoretically (Allard et al 2001; Chabrier & Baraffe 2000; Burrows et al. 2001) and observationally (Basri 2000; Kirkpatrick 2005). Most field BDs have been discovered in wide-field optical and near-infrared surveys such as the Two Micron All-Sky Survey (2MASS; Skrutskie et al. 1997), the Deep Near-Infrared

Southern Sky Survey (DENIS; Epchtein et al. 1997), and the Sloan Digital Sky Survey (SDSS; York et al. 2000). Until recently, these efforts have centered primarily on optical and near-infrared spectroscopy (Kirkpatrick et al. 1991; Geballe et al. 2002; Leggett et al. 2002; McLean et al. 2003), owing to the extreme difficulty of obtaining mid-infrared data from the ground (Matthews et al. 1996; Creech-Eakman et al. 2004, Sterzik et al. 2005). Recent observations by Cushing et al. (2005) have extended ground-based spectroscopy of BDs out to ~4 μm. The launch of the *Spitzer Space Telescope* (Werner et al. 2004) has allowed an unprecedented look at BD atmospheres using the Infrared Array Camera (IRAC, Fazio et al. 2004) and Infrared Spectrograph (IRS, Houck et al. 2004) instruments. Patten et al. (2006) have used the *Spitzer* IRAC instrument to obtain mid-infrared photometry of M, L, and T dwarfs; Cushing et al. (2006) and Roellig et al. (2004) have performed low resolution spectroscopy of M, L, and T dwarfs using the IRS.

Although the peak flux of these objects is centered near $\lambda$ ~1 μm, the mid-infrared offers useful opportunities for understanding fundamental properties of BDs. The fundamental bands of many molecular species common in the atmospheres of BDs, including $H_2O$, $NH_3$, and $CH_4$, typically occur in the mid-infrared, where their opacities are more accurately characterized. Signatures of non-equilibrium chemistry as well as silicate, iron, and corundum cloud structures should be apparent in this wavelength regime (Saumon et al. 2003a, b). In addition, R=600 spectroscopy offers the opportunity to test whether or not trace molecular species such as $CO_2$, $C_2H_2$, $C_2H_4$, $C_2H_6$, and HCN are detectable in multiple systems, perhaps even in emission, as found in our own solar system's giant planets (Orton et al. 2006, Saumon et al. 2003b, Yelle 2000). Emission lines of low abundance, but high absorption cross section, species would indicate a temperature inversion or stratosphere, perhaps arising from absorption of a primary star's UV radiation or even dynamical heating. Moderate resolution mid-infrared spectroscopy can potentially allow detailed tests of these ideas.

The low resolution (R~90) mid-infrared BD spectral sequence of Roellig et al. (2004) and Cushing et al. (2006) can be expanded using the moderate resolution R=600 capability of the *Spitzer* IRS. In this paper, we present IRS R=600 observations from 10 – 19 μm of M, L, and T dwarfs. Section 2 describes the observations, data reduction, and absolute flux calibration of the spectra, and Section 3 presents a preliminary comparison of the spectra to model atmospheres. Conclusions are given in Section 4.

## 2. Observations and Data Reduction

Our sample for the IRS moderate resolution observations consists of 13 M, one L, and two T dwarfs observed as part of the "Dim Suns" IRS Science Team Guaranteed Time Observer (GTO) program[1] (Roellig et al. 2004). These objects were drawn from the larger pool of Dim Suns objects (Cushing et al. 2006). They were selected to be sufficiently bright at R=600. Observations for additional comparable objects could not be used since they failed for various reasons. The IRS is composed of four modules capable of performing low- ($R = \lambda/\Delta\lambda \sim 90$) to moderate-resolution (R = 600) spectroscopy from 5.3 to 38 μm. We used the Short-High (SH) module that covers from

[1] Databases of known L and T dwarfs can be found at http://www.dwarfarchives.org and http://www.jach.hawaii.edu/skl/LTdata.html

10 to 19 μm at R = 600 in ten orders. A log of the observations, including the *Spitzer* AOR key, spectroscopic mode, and total on-source integration time is given in Table 1. Although both optical and infrared spectral types are listed in Table 1, we will hereafter use optical types for the M and L dwarfs (Kirkpatrick et al. 1991, 1999) and infrared types for the T dwarfs (Burgasser et al. 2006) unless otherwise noted. In addition, we hereafter abbreviate the numerical portions of the 2MASS, SDSS, and DENIS target designations as Jhhmm±ddmm, where the suffix is the sexigesimal Right Ascension (hours and minutes) and declination (degrees and arcminutes) at J2000 equinox.

The observations consisted of a series of exposures obtained at two positions along each slit. The raw IRS data were processed with the IRS pipeline (version S14) at the *Spitzer* Science Center. No dedicated off-sky observations were taken for any of the objects except 2MASS J0559-1404 and Gl 65AB. For these two objects, separate off-source observations of the sky were taken with integration time equal to that of the on-source observations. One of our observations of ε Indi Ba/Bb failed due to a bad peak-up, and we ended up with an observation of blank sky. This failed observation was used as a surrogate sky observation for the remaining objects in Table 1. A more detailed discussion of the Zodiacal background subtraction methodology is given below.

Post-launch, the IRS arrays were discovered to suffer from rogue pixels caused by radiation damage. A rogue pixel is a pixel with abnormally high dark current and/or photon responsivity (a hot pixel) that manifests as pattern noise in an IRS image. The term "rogue" was used originally to distinguish pixels whose responsivity was unpredictable, but now the definition of rogue pixels includes those that are permanently as well as temporarily hot. Approximately 0.2% of all pixels in the Short-High array are rogue pixels, a fraction that is increasing with time as the arrays are irradiated throughout the *Spitzer* mission. The amplitude of the typical rogue pixel is many times larger than the relatively weak signals from the faint brown dwarf targets. The *Spitzer* Science Center (SSC) has provided a set of rogue pixel masks for each IRS campaign; these rogue pixels were identified by having dark count rates greater than 4-sigma above the mean dark current over the entire array. These masks were augmented with additional lower-level rogue pixels found through a careful visual examination of each object's coadded nods. The rogue pixels were masked using the *irsclean_mask* routine provided by the SSC.[2] This routine replaces the rogue pixel with an average over the remaining pixels in the spatial direction. It is important to remove rogue pixels from the data, since their effect is to create a sharp spike in the extracted spectrum, mimicking emission-like features.

The data were reduced using the Spectroscopic Modeling Analysis and Reduction tool (SMART; Higdon et al. 2004) written for the IRS. In the case of 2MASS J0559-1404 and Gl 65AB, where dedicated sky observations had been taken, the sky observations were subtracted from the on-source images. Separate sky observations taken at the same time as the science target are beneficial for two reasons: first, the rogue pixels appear in both image and sky frames at approximately the same value, allowing the excess pixel current to largely subtract away. Second, the contribution of the mid-infrared sky background is considerable compared to the BD flux, especially at longer wavelengths. It is necessary to subtract this background level; measuring it directly adjacent to the object eliminates uncertainties in the absolute value of the sky level. For the remainder of the objects, where no dedicated sky measurement was made, no sky was subtracted at this stage of the processing. All spectra were extracted using the full six pixel-wide aperture and wavelength calibrated using SMART. The edges of each spectral order were trimmed using an IDL routine.

---

[2] The software package irsclean_mask can be found at
http://ssc.*Spitzer*.caltech.edu/archanaly/contributed/irsclean/

For all objects except 2MASS J0559-1404 and Gl 65AB, the missed pointing on ε Indi Ba/Bb was used as a substitute sky observation. A spectrum was extracted from this missed pointing in the same manner as the other targets. Before this Zodiacal spectrum could be subtracted from the BD spectra, it had to be scaled to match the Zodiacal emission level in the direction of the BD target. Since all of these Short-High observations were taken in conjunction with observations using the Short-Low module, the flux seen in the Red and Blue Peakup cameras in the Short-Low module provided a convenient way to determine the relative Zodiacal emission levels. The ratios of the mean backgrounds in the red and blue peakup arrays between ε Indi Ba/Bb and the target were computed. The missed ε Indi Ba/Bb pointing was multiplied by these ratios to approximate the difference in zodiacal background levels between ε Indi Ba/Bb and other targets. Following extraction of the target spectrum, this substitute sky spectrum was subtracted from the target. Sky subtraction was performed after extraction since the substitute sky was observed at a significantly different time than each of the science targets, therefore requiring a different rogue pixel mask than each of the BDs. The pitfalls of this technique lie in the fact that the sky observations are not contemporaneous spatially or temporally with the target, and the Zodiacal background's spectrum varies substantially over the sky . Future IRS Short-High users have been advised by the *Spitzer* Science Center to collect contemporaneous sky observations to avoid this problem.

Observations of the standard star ξ Dra (K2 III) obtained as part of the normal IRS calibration observations were used to remove the instrument relative spectral response function (RSRF) and flux calibrate the science targets. The standard star spectra were extracted in a similar fashion to the science targets. A model spectrum from Cohen et al. (2003) was used to remove the intrinsic stellar spectral energy distribution from the raw standard star spectra. This technique also serves as a first-order correction to fringes caused by interference within the active layer of the detector array. An RSRF was calculated for each of the two nod positions and was applied to each science target's two nods separately. We attempted to remove the remaining fringe signal from our ξ Dra observations using the IRSFRINGE routine provided as part of the SMART data reduction package. According to the *Spitzer* Science Center IRSFRINGE documentation, the fringe amplitude is typically ~2-5%. No further attempt at "de-fringing" the BD data was made, since in BDs, $H_2O$ bands create a complicated pattern of absorption features that could be confused with actual fringes caused by interference in the IRS. Short-High absolute fluxes were compared with the low resolution data of Cushing et al. (2006) and match to within 7-8% or better unless otherwise noted. All Short-High data were scaled to match the absolute fluxes of the low resolution data, which were calibrated using the IRAC Band 4 fluxes of Patten et al. (2006).

## 3. Discussion

### 3.1 M Dwarfs

In M dwarfs (2400 K ≤ $T_{eff}$ ≤ 3800 K), C, N, and O are found primarily in CO, $N_2$, and $H_2O$. The spectral morphology of M dwarfs is therefore shaped primarily by $H_2O$ absorption bands, although TiO and VO bands dominate in the optical. Since $N_2$ is a homonuclear molecule, it cannot radiate in the dipole approximation and therefore shows no spectral signatures in M and L dwarfs. The primary feature of interest in the mid-infrared spectra of M dwarfs is the "break" due to the $\nu_2$ fundamental absorption feature of $H_2O$ at ~6.5 μm. Since the shortest wavelength observable with the IRS Short-High module is 10 μm, this feature is unobservable in our R=600 spectrum. The M dwarf spectra appear featureless in the Short High data. Figure 1 depicts the Short-High spectra of our M dwarf sample; the signal to noise ratio is typically ~100 for these objects. The dominant error for these high-SNR observations is fringing, ~2-5% in amplitude.

### 3.2 L Dwarf - DENIS J0255-4700

Figure 2 depicts the sole L dwarf in our sample, the L8 dwarf DENIS J0255-4700 (Martín et al. 1999, Kirkpatrick et al. 2000). In L dwarf spectra, $H_2O$ and CO absorption strengthens. The mid-L dwarfs also show the onset of the $\nu_3$ fundamental band of $CH_4$ centered at ~3.3; however, this feature is not observable in the IRS Short-High module. Although the abundance ratio $CO/CH_4$ = 1 for T ~ 1100 K and P=1 bar (Lodders & Fegley 2002), the upper atmospheres of mid-L dwarfs can be cooler than 1100 K even if $T_{eff}$~1400 K, resulting in the appearance of $CH_4$ in the emergent spectrum. While the 10 – 20 μm spectral region does not contain any strong $CH_4$ features, $CH_4$ is an important absorber at shorter wavelengths and affects the spectral energy distribution via flux redistribution.

As spectral type progresses from L to T, a new molecular feature appears, the 10.5 μm Q-branch band of $NH_3$ at ~1200 K. $NH_3$ becomes the nitrogen-bearing gas species that dominates the spectrum between 10-14 μm both because it becomes more abundant ($NH_3/N_2$ > 1 for T < 700 K at P=1 bar - Lodders & Fegley 2002) and since $N_2$ has no observable features in the mid-infrared. These $NH_3$ features mix with $H_2O$ features, which has a complex band structure throughout the 10 – 19 μm coverage of the IRS-SH module. Figure 2 shows the comparison of the 10 – 12 μm region of our R=600 observation of DENIS J0255-4700 to a cloudy model by Marley et al. (2002); the model has been normalized to the data. We observed DENIS J0255-4700 with SNR ~30 at 10 μm and ~16 at 19 μm. A composite spectrum for this object extending from the optical through the mid-infrared is best matched by an equilibrium model with $T_{eff}$=1400 K and log g = 5.0 (M. Cushing, priv. comm.), if the model's absolute flux is uncalibrated. The object's parallax has been measured (Costa et al. 2006); when we adjust the model flux for the measured distance, we predict an absolute flux at 10 μm of ~14 mJy, significantly higher than the observed 9.4 ± 0.5 mJy. This may be due to the lack of reliable age and gravity measurements for this object, as well as inherent uncertainties in the models;

binarity and metallicity variations may also play a role. Also, we are unable to establish the presence of $NH_3$ in this spectrum. Models indicate that at this effective temperature, a SNR > 5000 would be required to detect the strongest $NH_3$ feature at 10.4 μm. Nevertheless, the models match the individual spectral features (due to $H_2O$ opacity) of this object remarkably well once it is normalized.

Cushing et al. (2006) define an $NH_3$ index versus spectral type; however, since it is defined as the mean flux of a 0.3 μm window around 10.0 μm, it is not possible to precisely compare this index with the R=600 data presented here, since the IRS Short-High module cuts on at 10.0 μm. The Short-High results are in reasonably good agreement with the Short-Low data, although the systematic offset beyond 12 μm may be due to the lack of a nearby and contemporaneous sky measurement for this object, as described in Section 2, or problems with the model. As there is no Short-Low data longward of 15 μm, we cannot compare with Short-Low; however, at shorter wavelengths, the Short-Low and Short-High data match reasonably well.

### 3.3 T Dwarfs – ε Indi Ba/Bb and 2MASS J0559-1404[3]

The $CH_4$ feature that appears at $T_{eff}$ ~ 1600 K becomes very strong in T dwarfs, dominating the spectrum between 7 – 9.2 μm (Cushing et al. 2006; Roellig et al. 2004). The appearance of $CH_4$ overtone and combination bands in the near-infrared signals the transition to the T spectral class (600 ≤ $T_{eff}$ ≤ 1400 K). The Si-, Fe-, and Al-bearing condensates that play a pivotal role in shaping the emergent spectrum of L dwarfs become less critical as they sink below the observable photosphere in T dwarfs, resulting in a largely cloud-free atmosphere in mid to late T dwarfs. The 10.5 μm band of $NH_3$ overlaps with absorption features of $H_2O$ between 9 – 16 μm as the temperature decreases. Beyond 15 μm, almost all features are due to $H_2O$, with the exception of weak $NH_3$ bands between 32 – 46 μm for models with $T_{eff}$ < 800 K (Saumon et al. 2003b).

We observed two T dwarfs in our sample, the binary T1/T6 system ε Indi Ba/Bb (Scholz et al. 2003), and the T4.5 dwarf 2MASS J0559-1404 (Burgasser et al. 2000).

**ε Indi Ba/Bb:** The ε Indi system consists of three members: a primary K4.5V star (ε Indi A) and a close pair consisting of two T dwarfs separated by 0.732 arcsec (ε Indi Ba/Bb; McCaughrean et al. 2004), located 400 arcsec away from the primary. The IRS does not have the spatial resolution to resolve ε Indi Ba and Bb, so the measured spectrum is the sum of the two individual spectra. Since in the near-infrared ε Indi Ba/Bb is spatially resolved, bright, and very well-characterized, and since the metallicity, age, and distance of the primary star (ε Indi A) are known, the physical parameters of the Ba and Bb components are very well constrained. The bolometric luminosity of each component (McCaughrean et al. 2004), the age of the system (Lachaume et al. 1999) and the metallicity of ε Indi A provide the constraints required to obtain $T_{eff}$ and the gravity from the brown dwarf evolution models. The near-infrared data indicate that ε Indi Ba has a spectral type of T1, and ε Indi Bb has a spectral type of T6 (McCaughrean et al. 2004).

[3] These sources are also referred to as ε Indi BC.

The spectra of late T dwarfs are best modeled with cloudless atmospheres, and we find $T_{eff}$=840K and log g=4.89 for the Bb component using an evolution sequence for cloudless brown dwarfs (Roellig et al. 2004). A T1 spectral type falls in the L-T transition where the role of clouds is still unclear, so we consider both cloudy ($f_{sed}$ = 3) and cloudless models for the Ba component, giving $T_{eff}$=1210K and log g=5.10 and $T_{eff}$=1250K and log g=5.13, respectively. Typical uncertainties are < 90K on $T_{eff}$ and ~ 0.15 on log g. With these parameters, the T1 component is about twice as bright in the mid-infrared as the T6 component. The 10-11 μm band of $NH_3$ is very strong in T6 and essentially absent in T1 dwarfs (Cushing et al. 2006). Nevertheless, the strong $NH_3$ band in the T6 is easily detected in the combined IRS-SH spectrum (Fig. 3).

We analyze the $NH_3$ band in the spectrum of ε Indi Ba/Bb by computing composite model spectra for 6 different cases, considering that the Ba component may or may not be cloudy, and also including the possibility that the $NH_3$ abundance may be reduced by vertical transport in the atmosphere of both components. Vertical transport in the atmosphere can lead to non-equilibrium abundances for slowly reacting species such as CO and $N_2$, with the net effect of enhancing the upper atmosphere abundances of CO and $N_2$ to the detriment of $CH_4$, $H_2O$, and $NH_3$ (Saumon et al. 2003; Fegley & Lodders 1996; Griffith & Yelle 1999; Lodders & Fegley 2002). Evidence for non-equilibrium abundances of $NH_3$ has been detected in the late T dwarfs Gl 229B and Gl 570D (Saumon et al 2000; 2003; 2006). The efficiency of vertical transport is parametrized by the vertical eddy diffusion parameter, $K_{zz}$ (Griffith & Yelle 1999). We consider models with $K_{zz}$=0 (equilibrium), $10^2$ and $10^4$ $cm^2/s$. The six combinations of parameters are given in Table 2. Cloudy atmospheres are parametrized by a sedimentation parameter, $f_{sed}$, (Ackerman & Marley 2001) which we choose to be $f_{sed}=3$, a value we found suitable to reproduce the near infrared colors of early T dwarfs (Marley et al, in prep.). Cloudless models have $f_{sed}$ = infinity. Given the effective temperature and gravity of each component, the evolution sequences provide their radii. The trigonometric parallax (Perryman et al. 1997) then allows us to compute the absolute composite model flux at Earth to compare with the IRS spectrum, *without any arbitrary renormalization* (Roellig et al. 2004). Figure 3 depicts models A and C from Table 2 compared to our observed R = 600 spectrum of ε Indi Ba/Bb. The models were smoothed to the IRS Short-High point source spectral resolution of $\lambda/\Delta\lambda$ = 600 using SMART and then binned to the instrument's wavelength sampling.

Figure 4 shows all six models compared to the $NH_3$ spectral region of ε Indi Ba/Bb. Our observation yielded SNR ~50 at 10 μm and ~15 at 19 μm, so the major uncertainty is the overall flux calibration of the Short-Low data to which we scaled the Short-High data. The Short-Low observation of ε Indi Ba/Bb has a ±5% flux calibration error (Cushing et al. 2006, Patten et al. 2006). The effects of non-equilibrium chemistry can be seen in the depletion of $NH_3$ relative to the equilibrium models (A & B). The goodness of fit of the six model spectra is determined by computing the $\chi^2$ between the models and the data. The uncertainty ($\sigma$) on the fitted $\chi^2$ is obtained by fitting 5000 simulated data sets generated by adding a random Gaussian noise distribution to each pixel. The Gaussian's width is given by the individual pixel's error bar. The $\chi^2$ fit reveals that the best-fit model is a cloudless non-equilibrium atmosphere with $K_{zz}$ = $10^2$ $cm^2/s$ for both components, model C. Note that both values of $K_{zz}$ used in non-equilibrium models lead to nearly identical mid-infrared spectra as the $NH_3$ depletion is rather insensitive to the mixing efficiency in late T dwarfs (Saumon et al. 2006). Therefore, model D and model C give equally good fits to the data. Model C is 27-$\sigma$ better than the best equilibrium model, A. Model C still fits reasonably well (3-$\sigma$) even when the data are scaled down by the full 5% flux calibration error. Even though we cannot obtain individual spectra of $\varepsilon$ Indi Ba and Bb with the IRS, each component's parameters are so well-constrained that we find this good evidence of non-equilibrium chemistry in $\varepsilon$ Indi Bb. The T1 component may be similarly affected, but its higher temperature results in a low $NH_3$ abundance even in equilibrium, and any non-equilibrium effect would be much less in the hotter component. $\varepsilon$ Indi Bb joins the ranks of Gl 229B and Gl 570D in a growing set of T dwarfs where non-equilibrium $NH_3$ abundances have been observed.

**2MASS J0559-1404:** 2MASS J0559-1404 is a T4.5 dwarf (Burgasser et al. 2006). Golimowski et al. (2004) find $T_{eff}$ ~1425 K if 2MASS J0559-1404 is a single 3 Gyr old dwarf; if it is a multiple system, the $T_{eff}$ will be lower (~1150 K for an equal mass pair). It is possible that 2MASS J0559-1404 is actually a multiple system based on its brightness (Burgasser et al. 2000), although efforts to spatially resolve it into multiple components using the *Hubble Space Telescope* have thus far failed (Burgasser et al. 2003). By $T_{eff}$ ~1400 K, the condensate clouds have sunk to altitudes that are well below the 10 μm photosphere, so they should not directly affect the emergent mid-infrared spectrum of 2MASS J0559-1404. With a T4.5 spectral type, we expect minimal contribution from clouds and the strongest $NH_3$ band in our sample. The SNR of the Short-High observation (~14 at 10 μm and 3.5 at 19 μm) is lower than that of $\varepsilon$ Indi Ba/Bb owing to its greater distance, intrinsic faintness, and slightly higher sky background. Regardless, we clearly observe the $NH_3$ band in its spectrum (Figure 5).

The equilibrium model that best fits a composite spectrum extending from optical through mid-infrared wavelengths indicates that $T_{eff}$ = 1100-1200 K, log g = 5.0 – 5.5, with a cloud sedimentation parameter $f_{sed}$ = 4. (M. Cushing, priv. comm.) We find that models with these parameters, assuming that it is a single dwarf and using its known parallax (Dahn et al. 2002), predict a 10 μm flux between 1.3 – 2.3 mJy, compared to the measured 10 μm flux of 2.7±0.2 mJy. Although this may support the notion that this is an unresolved multiple system, this may simply reflect the inherent uncertainties in the models for this object and the lack of a reliable age/gravity. However, we can renormalize the model fluxes to fit only the shape of the spectrum in the region between 10 – 11 μm, which is where differences in the four models are most prominent. We find that of the four equilibrium models studied ($T_{eff}$ = 1100 K and 1200 K, log g = 5.0 and 5.5), we can only say that the model with $T_{eff}$ = 1100 K and log g = 5.5 provides the worst fit at the 6-$\sigma$ level compared with the other three models, which all had similar $\chi^2$ values within the error bars. When scaled down by the full 5% flux calibration uncertainty, this drops to 3.5-$\sigma$.

We attempted to test whether or not non-equilibrium chemistry is apparent in the 10 - 11 μm $NH_3$ feature. Fitting the equilibrium and non-equilibrium ($T_{eff}$ = 1100, log g = 5.0, $K_{zz}$ = $10^2$ $cm^2/s$) models to the 10 – 11 μm region of the Short-High spectrum after normalizing by the flux showed that the non-equilibrium model is preferred at the 2-σ level, but this distinction disappears after scaling down by the full ±5% flux calibration error (Figure 6). Nonetheless, the models fit the numerous absorption features of $NH_3$ and $H_2O$ well.

### 3.4 Trace Molecular Species

In L and T dwarf atmospheres, $H_2O$ remains in the gas phase and in the presence of sufficient UV flux can participate in photochemistry, forming such species as HCO and $CH_2O$ (Friedson, Wilson & Moses 2003). For example, solar fluorescence causes both CO and HCN to be observed in emission in the mid-infrared on Neptune, even though it is 30 AU from the Sun, as well as Uranus (Orton et al. 2006; Rosenqvist et al. 1992). In the absence of a stratosphere, these molecules would appear in absorption. The equilibrium abundances of these molecules should be quite low in the absence of photochemistry. Saumon et al. (2003b) calculate the minimum enhancement factor (ε) of the abundance of several trace species required for detection. The minimum enhancement factor ε is defined as the ratio of the column density of the trace species above the photosphere that is needed for a detection to the chemical equilibrium column density above the photosphere. Table 1 in Saumon et al. (2003b) gives the wavelength of the strongest opacity feature, the $T_{eff}$ that is most favorable for detection, and the enhancement factor. They find that only $CO_2$ is likely to be detected in absorption, particularly at higher metallicity and lower gravity.

Figure 7 depicts the cross-sections of various trace molecular species that might be expected to occur for log g = 5 and $T_{eff}$ = 1300 K: $CO_2$, $C_2H_2$, $C_2H_4$, $C_2H_6$, and HCN (Saumon et al. 2003b). Comparison of these opacities to our data show no signatures of any of these species, although only $CO_2$ was expected to appear at 15 μm in absorption in an equilibrium model with $T_{eff}$ = 1200 K and log g = 4. We also considered the effects of LiCl (Weck et al. 2004) and $H_2$-$N_2$ (L. Frommhold, priv. comm.); neither of these molecules are expected to have a significant effect on any of the objects we observed compared to $NH_3$ and $H_2O$. This indicates that, at least at this spectral resolution and sensitivity, trace molecular species have little influence on the emergent spectra of DENIS J0255, 2MASS J0559-1404, and ε Indi Ba/Bb.

We observed a number of binary brown dwarf systems. In some cases, we observed the system's hotter component, while in others, we observed only the cooler component; in one case, we observed both. In the case of Gl 65A/B, we observed both components simultaneously, since with an angular separation of 2 arcsec, Gl 65A and Gl 65B are unresolved by the IRS, which has a beam size of ~2.9 arcsec at 10 μm. The two M dwarfs, an M5.5 and an M6 dwarf (Kirkpatrick et al. 1991), are separated by only 5.1 AU. As shown in Figure 1, no emission lines were detected in our observation of this system, which has the smallest separation between companions in our sample (see Table 3), nor in any of the other binary M dwarfs we surveyed. In another case, we observed a system's secondary: the M8 dwarf VB 10 (Gl 752B), which is separated by ~522 AU from its companion, the M2.5 dwarf star Gl 752A (Kirkpatrick et al. 1991). Figure 1 shows that its spectrum is also unremarkable. In the case of Gl 229A, we are observing the system's hotter component, and its spectrum reveals no emission features. Fringing effects (as noted above) constitute the dominant error source for these high signal to noise observations. Since the fringe signal is ~2-5% of the continuum amplitude, a single-bin line would have to have exceed the continuum signal by ~25% to constitute a 5-$\sigma$ detection. Our IRS-SH data do not reveal any unusual spectral features in the mid-infrared spectra of M dwarfs, as is the case at the lower resolution of the IRS-SL module (Cushing et al. 2006).

## 4. Conclusions

We have presented R=600 spectra of M, L, and T dwarfs from 10 – 19 $\mu$m. These spectra have prominent absorption bands of $H_2O$ and $NH_3$ apparent in the IRS SH data. $H_2O$ absorption features are present throughout the MLT sequence (although $H_2O$ in M dwarfs is only seen at wavelengths not observable by the IRS SH module), while $NH_3$ appears in T dwarfs. Additionally, we find evidence for non-equilibrium chemistry in the atmosphere of the T6 component of the binary $\varepsilon$ Indi Ba/Bb, providing further support to the concept that vertical transport plays a significant role in T dwarf atmospheres. We found that models with no clouds, at least over the pressure ranges probed by the mid-infrared data, best fit the T1 dwarf $\varepsilon$ Indi Ba. We find no evidence of emission features, which would have indicated a stratosphere, in any of our targets. We see no evidence of trace molecular species in absorption.

## 5. Acknowledgements

We thank the entire team of the *Spitzer Space Telescope* for their hard work and dedication. This publication makes use of data from the Two Micron All Sky Survey, which is a joint project of the University of Massachusetts and the Infrared Processing and Analysis Center, and funded by the National Aeronautics and Space Administration and the National Science Foundation. This research has made use of the NASA/IPAC Infrared Science Archive, which is operated by the Jet Propulsion Laboratory, California Institute of Technology, under contract with NASA. This research has made use of the SIMBAD database, operated at CDS, Strasbourg, France, NASA’s Astrophysics Data System Bibliographic Services, and the M, L, and T dwarf compendium housed at DwarfArchives.org and maintained by Chris Gelino, Davy Kirkpatrick, and Adam Burgasser. This work was supported in part by the United States Department of Energy under contract W-7405-ENG-36. This work is based on observations made with the *Spitzer Space Telescope*, which is operated by the Jet Propulsion Laboratory, California Institute of Technology under NASA contract 1407. Support for this work was provided by NASA's Office of Space Science. A. Mainzer, T. Roellig and M. Marley would like to acknowledge the support of the NASA Science Mission Directorate.

## 6. References

Ackerman, A. S., Marley, M. S. 2001 ApJ 556, 872

Allard, F., Hauschildt, P. H., Alexander, D. R., Tamanai, A., & Schweitzer, A. 2001, ApJ, 556, 357

Basri, G., Marcy, G. W., & Graham, J. R., 1996, ApJ, 458, 600

Basri, G. 2000, ARA&A, 38, 485

Becklin, E. E., Zuckerman, B. 1988 Nature 336, 656

Burgasser, A. J., Wilson, J. C., Kirkpatrick, J. Davy, Skrutskie, M. F., Colonno, M. R., Enos, A. T., Smith, J. D., Henderson, C. P., Gizis, J. E., Brown, M. E., Houck, J. R. 2000 AJ 120, 1100

Burgasser, A. J., Kirkpatrick, J. Davy, Brown, M. E., Reid, I. N., Burrows, A., Liebert, J., Matthews, K., Gizis, J. E., Dahn, C. C., Monet, D. G., Cutri, R. M., Skrutskie, M. F. 2002 ApJ 564, 421

Burgasser A.J., Kirkpatrick, J.D., Reid, I.N., Brown, M.E., Miskey, C.L., Gizis, J. E. 2003 ApJ 586, 113

Burgasser, A. J., Kirkpatrick, J. Davy, Liebert, J., Burrows, A. 2003 ApJ 594, 510

Burgasser, A. J., Geballe, T. R., Leggett, S. K., Kirkpatrick, J. Davy, Golimowski, D. A. 2006 ApJ 637, 1067

Burrows, A., Hubbard, W. B., Lunine, J. I., & Liebert, J. 2001, Rev. Mod. Phys. 73, 719

Chabrier, G. & Baraffe, I. 2000, ARA&A, 38, 337

Cohen, M., Wheaton, W. A., Megeath, S. T. 2003 AJ 126, 1090

Costa, E., Méndez, R., Jao, W.-C., Henry, T., Subasavage, J., Ianna, P. 2006 AJ 132, 1234

Creech-Eakman, M. J., Orton, G. S., Serabyn, E., & Hayward, T. L. 2004, ApJ, 602, L129

Cushing, M. C., Rayner, J. T., & Vacca, W. D. 2005, ApJ, 623, 1115

Cushing, M. C., Roellig, T. L., Van Cleve, J. E., Sloan, G. C., Wilson, J. C., Saumon, D., Leggett, S. K., Marley, M. S., Cushing, M. C., Kirkpatrick, J. D., Mainzer, A. K., Houck, J. R. 2006, ApJ, 648, 614.

Dahn, C. C., et al. 2002 AJ 124, 1170

Epchtein, N., et al. 1997, Messenger, 87, 27

Fazio, G. G., et al. 2004, ApJS, 154, 10

Fegley, B. J., & Lodders, K. 1996, ApJ, 472, L37

Friedson, A. J., Wilson, E. & Moses, J. I. 2003, BAAS, 35, 944

Geballe, T. R., et al. 2002, ApJ 564, 466

Goldman, B.; et al. 1999 A&A 351, 5

Golimowski, D. A., et al. 2004, AJ, 127, 3516

Griffith, C., Yelle, R. 1999 ApJ 519, 85

Hawley, S. L., Gizis, J. E., & Reid, I. N. 1996, AJ, 112, 2799

Henry, T., Subasavage, J., Brown, M., Beaulieu, T., Jao, W.-C., Hambly, N. 2004 AJ 128, 2460

Higdon, S. J. U., Devost, D., Higdon, J. L., Brandl, B. R., Houck, J. R., Hall, P., Barry, D., Charmandaris, V., Smith, J. D. T., Sloan, G. C., & Green, J. 2004, PASP, 116, 975

Houck, J. R., et al. 2004, ApJS, 154, 18

Kirkpatrick, J. Davy, Henry, T. J., McCarthy, D. 1991 ApJS 77, 417

Kirkpatrick, J. D., Henry, T. J., & Simons, D. A. 1995, AJ, 109, 797

Kirkpatrick, J. D., et al. 1999, ApJ, 519, 802

Kirkpatrick, J. Davy; Reid, I. N.; Liebert, J.; Gizis, J. E.; Burgasser, A. J.; Monet, D. G.; Dahn, C. C.; Nelson, B.; Williams, Rik J. 2000 120, 447

Kirkpatrick, J. Davy; Dahn, C. C.; Monet, D. G.; Reid, I. N.; Gizis, J. E.; Liebert, J.; Burgasser, A. J. 2001 AJ 121, 3235

Kirkpatrick, J. Davy, 2005 ARA&A 43, 195

Lachaume, R., Dominik, C., Lanz, T., Habing, H. 1999 A&A 348, 897

Leggett, S. K, et al. 2002 ApJ 564, 452

Leggett, S. K., Allard, F., Dahn, C., Hauschildt, P. H., Kerr, T. H., & Rayner, J. 2000, ApJ, 535, 965

Lodders, K. & Fegley, B. 2002, Icarus, 155, 393

Marley, M. S., Seager, S., Saumon, D., Lodders, K., Ackerman, A. S., Freedman, R. S., & Fan, X. 2002, ApJ, 568, 335

Martín, E. L., Delfosse, X., Basri, G., Goldman, B., Forveille, T., Zapatero Osorio, M. R. 1999 AJ 118, 2466

Matthews, K., Nakajima, T., Kulkarni, S. R., & Oppenheimer, B. R. 1996, AJ, 112, 1678

McCaughrean, M. J., Close, L. M., Scholz, R.-D., Lenzen, R., Biller, B., Brandner, W., Hartung, M., Lodieu, N. 2004 A&A 413, 1029

McLean, I. S., McGovern, M. R., Burgasser, A. J., Kirkpatrick, J. Davy, Prato, L., Kim, S. S. 2003 ApJ 596, 561

Nakajima, T., Oppenheimer, B. R., Kulkarni, S. R., Golimowski, D. A., Matthews, K., Durrance, S. T. 1995 Nature 378, 463

Oppenheimer, B. R., Golimowski, D. A., Kulkarni, S. R., Matthews, K., Nakajima, T., Creech-Eakman, M. 2001 AJ 121, 2189

Orton, G., Burgdorf, M., Meadows, V., Van Cleve, J., Crisp, D., Stansberry, J., Atreya, S. 2006 DPS #37, #22.06

Patten, B., Stauffer, J., Burrows, A., Marengo, M., Hora, J. L., Luhman, K., Sonnett, S., Henry, T., Raghavan, D., Megeath, T., Liebert, J., Fazio, G. 2006 ApJ in press

Perryman, M. A. C., et al. 1997, A&A, 323, L49

Rebolo, R., Martin, E. L., Basri, G., Marcy, G. W., & Zapatero-Osorio, M. R. 1996, ApJ, 469, L53

Rebolo, R., Zapatero Osorio, M. R., Madruga, S., Bejar, V., Arribas, S., Licandro, J. 1998 Science 282, 1309

Roellig, T. L., Van Cleve, J. E., Sloan, G. C., Wilson, J. C., Saumon, D., Leggett, S. K., Marley, M. S., Cushing, M. C., Kirkpatrick, J. D., Mainzer, A. K., Houck, J. R. 2004, ApJS 154, 418

Rosenqvist, J., Lellouch, E., Romani, P. N., Paubert, G., Encrenaz, T. 1992 ApJ 392, L99

Saumon, D. Marley, M.S., Lodders, K. & Freedman, R.S 2003a, in Brown dwarfs, IAU Symposium 211, E.L. Martin, ed. p 345.

Saumon, D., Marley, M. S., Lodders, K. 2003b astro-ph/0310805

Scholz, R.-D., McCaughrean, M. J., Lodieu, N., Kuhlbrodt, B. 2003 A&A 398L, 29

Skrutskie, M. F., et al. 2006, AJ, 131, 1163

Sterzik, M. F., Pantin, E., Hartung, M., Huelamo, N., Kaufl, H. U., Kaufer, A., Melo, C., Nurnberger, D., Siebenmorgen, R., & Smette, A. 2005, A&A, 436, L39

Yelle, R. V., 2000, in From Giant Planets to Cool Stars, ASP Conf. Series Vol 212, C. A. Griffith & M. S. Marley, Eds. (ASP: San Francisco), 267.

York, D. G., et al. 2000, AJ, 120, 1579

Weck, P. F., Schweitzer, A., Kirby, K., Hauschildt, P. H., Stancil, P. C. 2004 ApJ 613, 567

Werner, M. W., et al. 2004 ApJS 154, 1

Table 1. Log of the IRS R=600 Observations.

| Object | Optical Sp. Type[a] | Infrared Sp. Type[a] | AOR Key | On-Source Exp. Time (sec) |
|---|---|---|---|---|
| Gl 229A[b] | M1 V | … | 4185856 | 63 |
| Gl 1 | M1.5 V | … | 3873792 | 63 |
| G 196-3A[b] | M2.5 V | … | 3879168 | 2886 |
| Gl 674 | M2.5 V | … | 3874304 | 63 |
| Gl 687 | M3 V | … | 3874560 | 38 |
| Gl 849 | M3.5 V | … | 3873024 | 188 |
| GJ 1001A[b] | M3.5 V | … | 4190464 | 1950 |
| Gl 65AB [b] | M5.5 V | … | 12246784 | 25 |
| GJ 1111 | M6.5 V | … | 3876096 | 964 |
| vB10 | M8 | … | 3876864 | 964 |
| DENIS J0255-4700 | L8 | … | 419558 | 1950 |
| ε Indi Ba/Bb [b] | … | T1 / T6 binary | 6313730 | 1950 |
| 2MASS J0559-1404 | T5 | T4.5 | 16202496 | 4148 |

[a] Optical spectral types of the M dwarfs are from Kirkpatrick et al (1991), Henry et al. (2004), Kirkpatrick et al. (1995), and Hawley et al. (1996). DENIS J0255-4700 optical spectral type from Kirkpatrick et al. (in prep, 2006). ε Indi Ba/Bb spectral type from Burgasser et al. (2006). 2MASS J0559-1404 spectral types from Burgasser et al. (2003) and Burgasser et al. (2006) .

[b] These objects are part of multiple systems.

Table 2: The six models compared to ε Indi Ba/Bb (Marley et al. 2002), where vertical transport is controlled by varying $K_{zz}$, the coefficient of vertical eddy diffusion; for cloud sedimentation efficiency ($f_{sed}$, Ackerman & Marley 2001), "nc" indicates cloudless models.

| Model | Ba | | | | | Bb | | | | | $\chi^2$ | $\Delta\chi^2$ |
|---|---|---|---|---|---|---|---|---|---|---|---|---|
| | $T_{eff}$ (K) | log g | $R/R_{sun}$ | $f_{sed}$ | $K_{zz}$ ($cm^2/s$) | $T_{eff}$ (K) | log g | $R/R_{sun}$ | $f_{sed}$ | $K_{zz}$ ($cm^2/s$) | | |
| A | 1250 | 5.13 | 0.0933 | nc | 0 | 840 | 4.89 | 0.0978 | nc | 0 | 26.90 | 0.35 |
| B | 1210 | 5.10 | 0.0987 | 3 | 0 | 840 | 4.89 | 0.0978 | nc | 0 | 21.79 | 0.30 |
| C | 1250 | 5.13 | 0.0933 | nc | $10^2$ | 840 | 4.89 | 0.0978 | nc | $10^2$ | 14.35 | 0.46 |
| D | 1250 | 5.13 | 0.0933 | nc | $10^4$ | 840 | 4.89 | 0.0978 | nc | $10^4$ | 14.11 | 0.53 |
| E | 1210 | 5.10 | 0.0987 | 3 | $10^2$ | 840 | 4.89 | 0.0978 | nc | $10^2$ | 48.16 | 0.24 |
| F | 1210 | 5.10 | 0.0987 | 3 | $10^4$ | 840 | 4.89 | 0.0978 | nc | $10^4$ | 63.14 | 0.24 |

Table 3. Multiple cool and ultracool dwarf systems observed. The components that were observed are highlighted in bold.

| Object Observed | Primary Sp. Type | Secondary Sp. Type | Separation from Primary (arcsec) | Separation from Primary (AU) |
|---|---|---|---|---|
| Gl 229A/B [a] | **M1 V** | T7p | 8 | 48.8 |
| G 196-3A [b] | **M2.5 V** | L2 | 16 | 300 |
| GJ1001A [e] | **M3.5 V** | L4.5 + L4.5 (GJ1001BC) | 18 | 174 |
| G 65 AB [c, d] | **M5.5 V** | **M6 V** | 2 | 5.1 |
| vB10 (Gl 752B) [b] | M2.5 V | **M8 V** | 3060 | 521.6 |
| ε Indi Ba/Bb [f] | K4.5 V | **T1 / T6** | 400 | 110 |

[a] Nakajima et al. 1995; Burgasser et al. 2002; Burgasser et al. 2006
[b] Kirkpatrick et al. 2001; Rebolo et al. 1998
[c] Kirkpatrick et al. 1991
[d] Oppenheimer et al. 2001
[e] Goldman et al. 1999
[f] McCaughrean et al. 2004

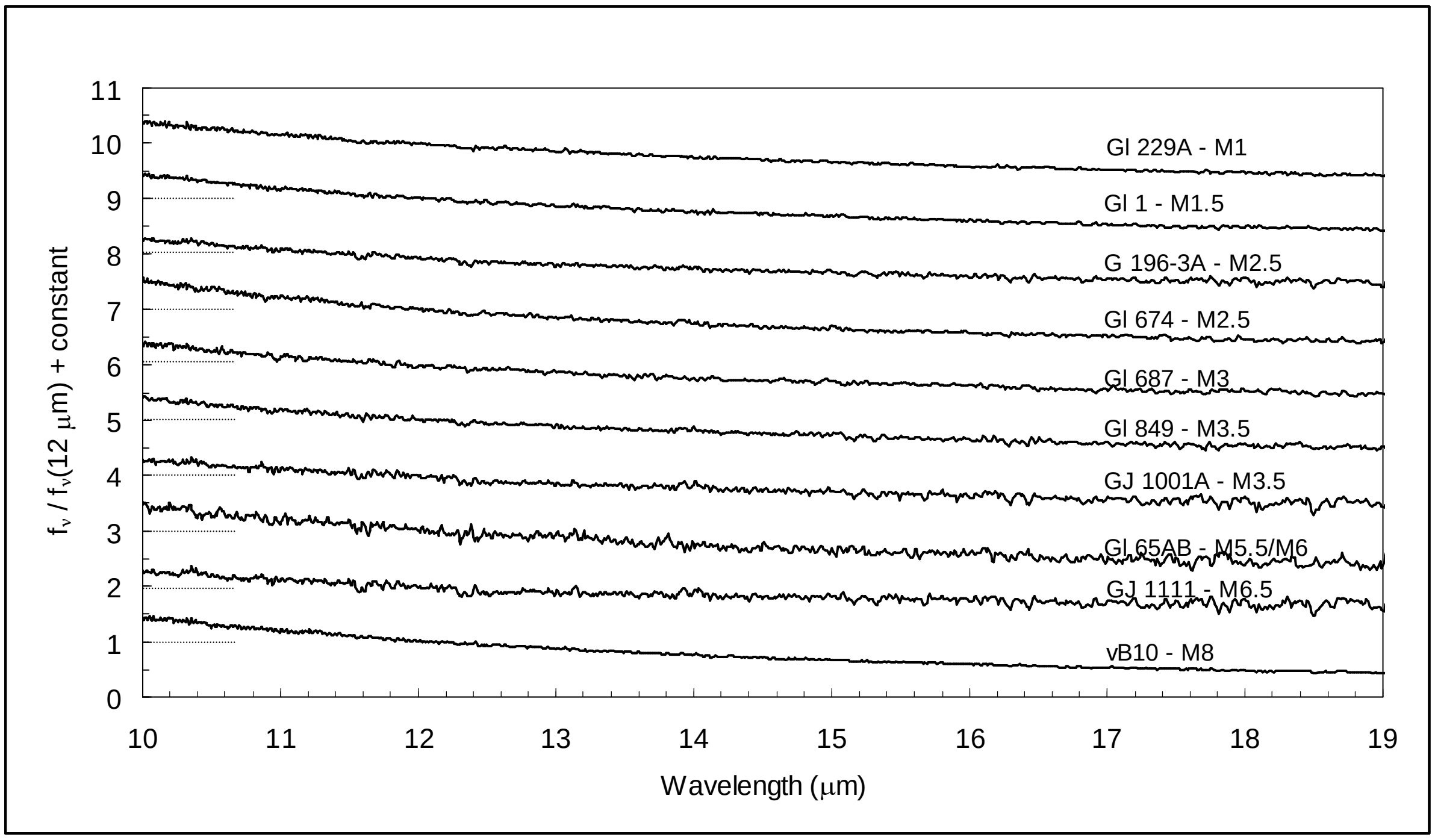

Figure 1. A spectral sequence of M dwarfs measured with the IRS Short-High module at R=600. The spectra have been normalized at 12 μm and offset by constants (dotted lines); the flux densities at 12 μm are 858 mJy (Gl229A), 620 mJy (Gl 1), 565 mJy (G 196-3A), 429 mJy (Gl 674), 649 mJy (Gl 687), 236 mJy (Gl 849), 452 mJy (GJ 1001A), 365 mJy (Gl 65AB), 842 mJy (GJ 1111), and 538 mJy (vB10).

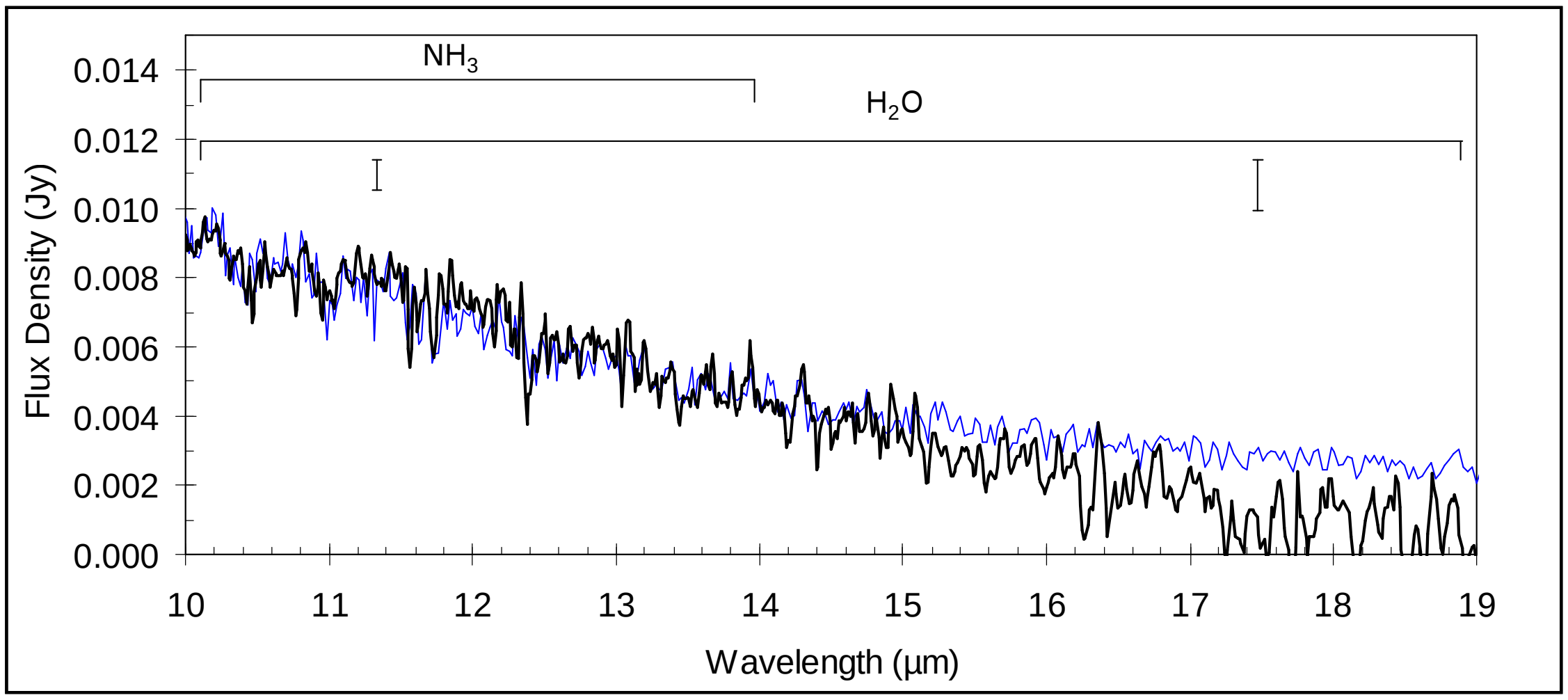

Figure 2. A cloudy model (Marley et al. 2002) with $T_{eff}$ = 1400 K, log g = 4.5, and sedimentation parameter $f_{sed}$ = 3, with $NH_3$ (thin blue line) compared to the L8 dwarf DENIS J0255-4700 Short-High data (heavy black line). Typical 1-σ error bars are shown; the uncertainty increases with wavelength. For this object, we did not observe a region of nearby sky. We used the sky from our observation of ε Indi Ba/Bb as an approximation; this may be responsible for the mismatch with the model at wavelengths longer than ~15 μm.

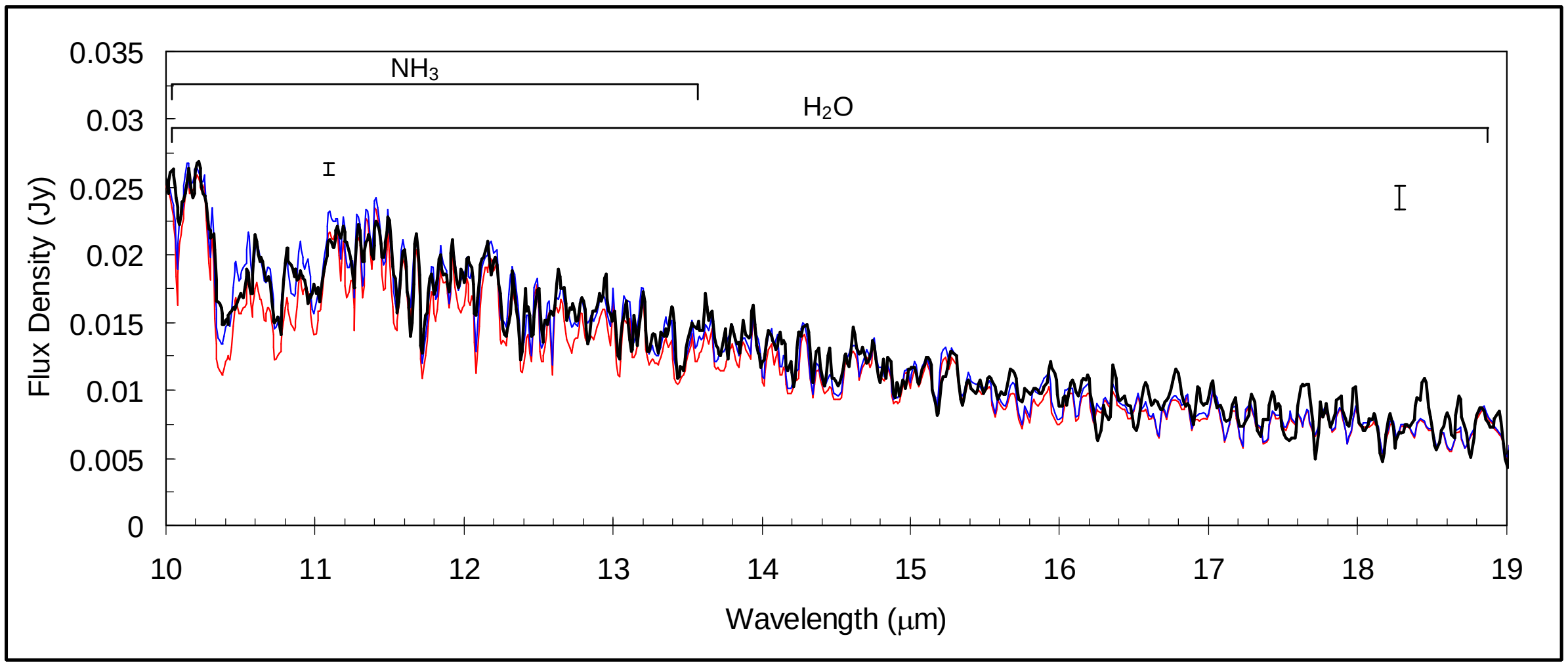

Figure 3: Comparison of the observed and modeled mid-infrared spectrum of ε Indi Ba/Bb. The data (heavy black curve) are compared to the equilibrium model A (thin red line, see Table 2) and the best fitting model, C (thin blue line). The models are plotted at spectral resolution of R=600. The typical 1-σ error is shown; the uncertainty increases with wavelength.

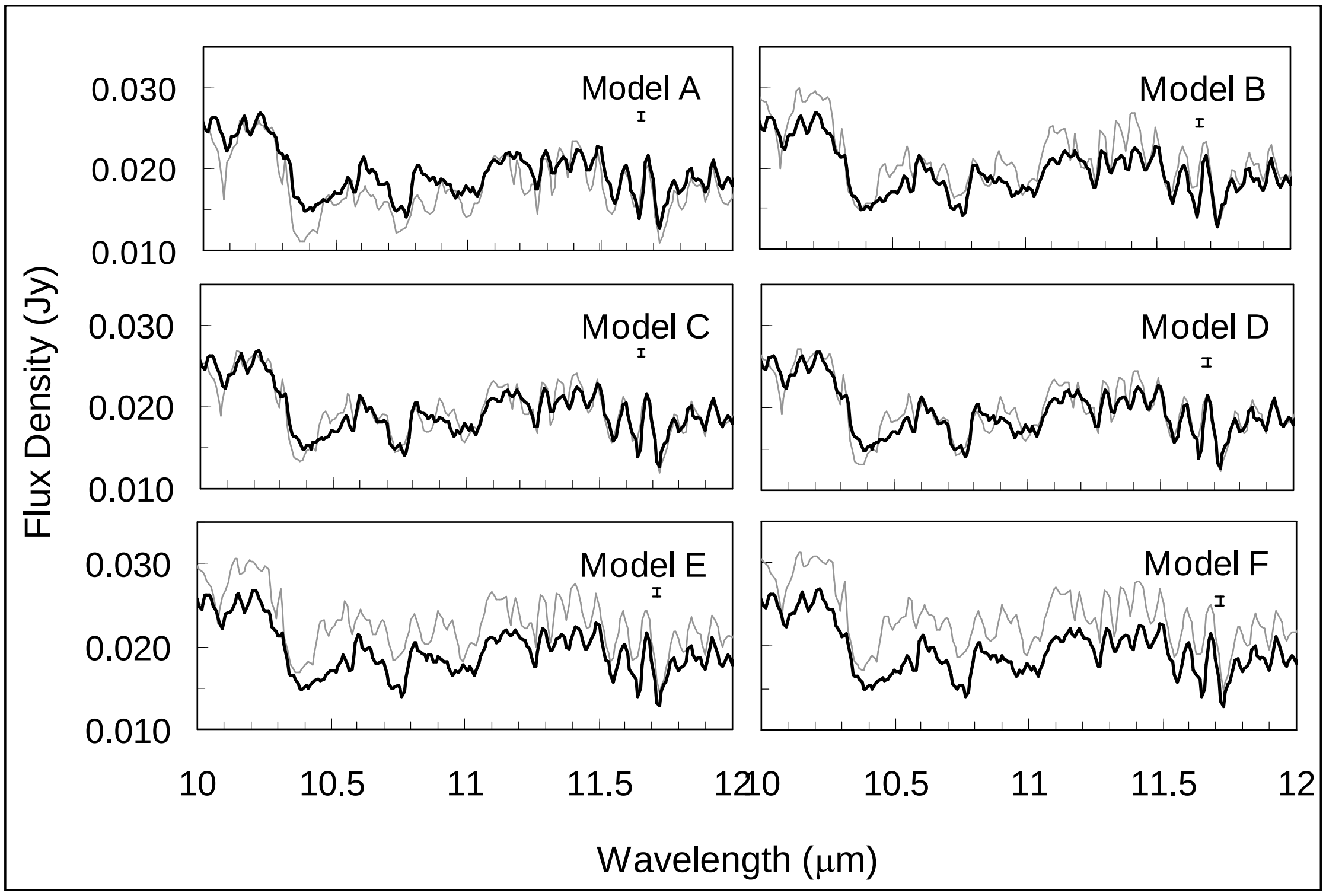

Figure 4: Comparison of the 10 – 12 μm region of the observed R=600 mid-infrared spectrum of ε Indi Ba/Bb to the six different models described in Table 2. Data: heavy black curve; models; thin gray line. The data are best fit by model C, a non-equilibrium cloudless model, which is a significantly better fit than model A, a cloudless equilibrium model. The typical 1-σ error bar is shown.

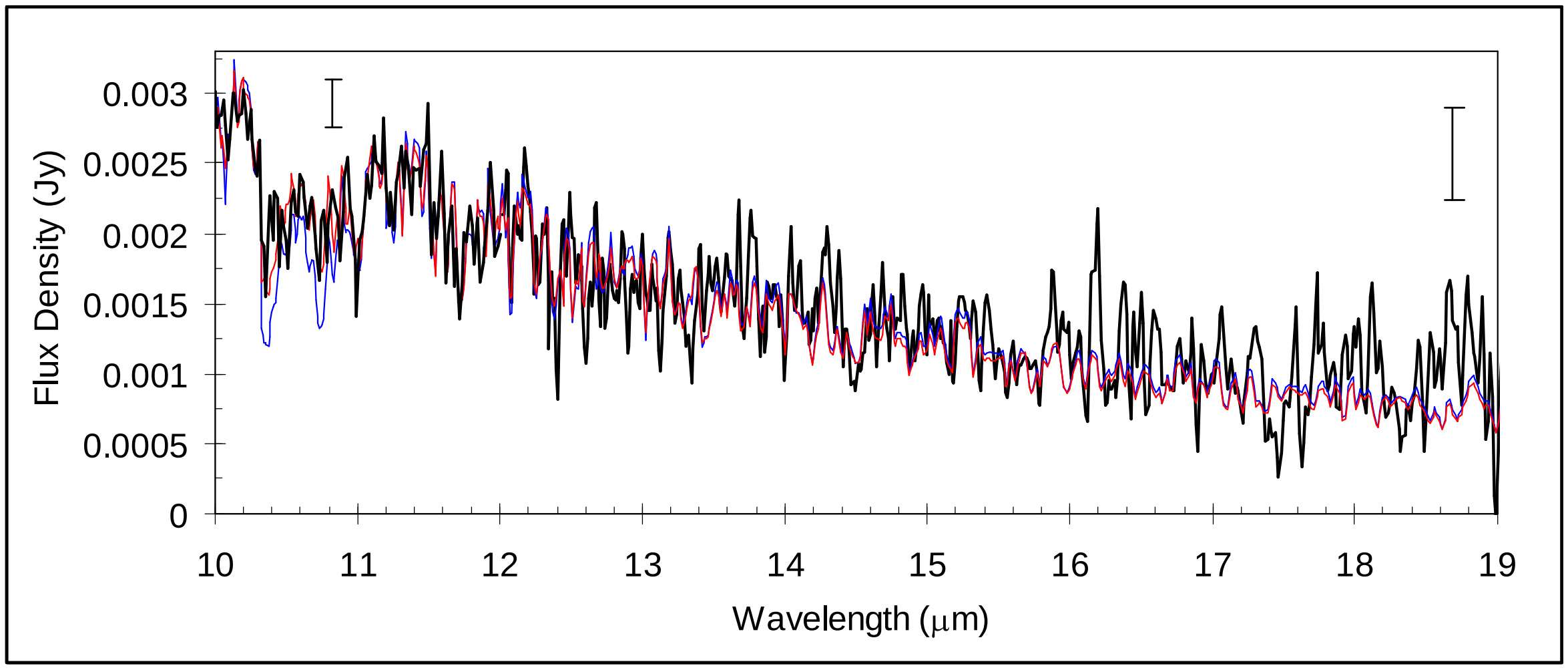

Figure 5: Comparison of cloudless models in chemical equilibrium (Marley et al. 2002) with $T_{eff}$ = 1100 K, log g = 5.477 (thin blue line) and $T_{eff}$ = 1200 K, log g = 5.477, (thin red line) to the IRS Short-High spectrum of 2MASS J0559-1404 (heavy black line). Both models have been normalized to the data. Typical 1-σ error bars are shown; the uncertainty increases with wavelength.

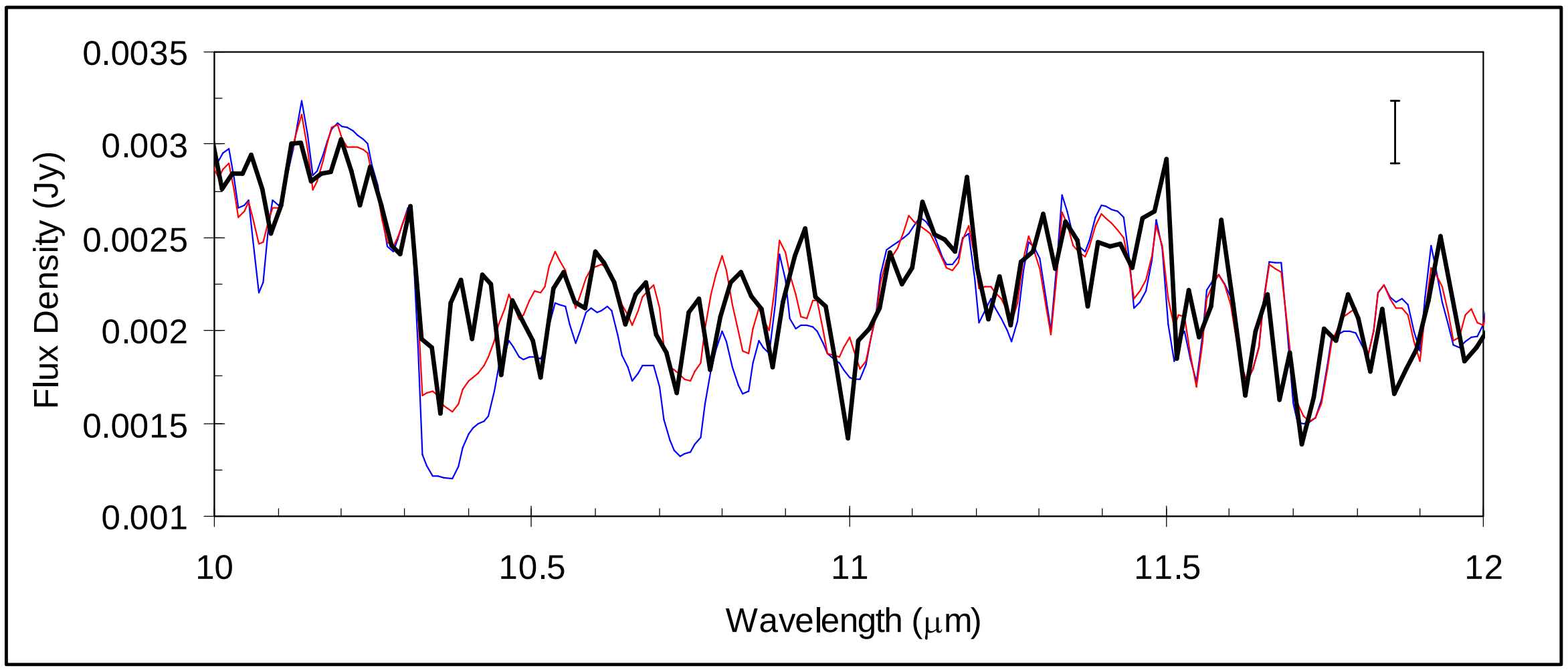

Figure 6: Detail of the 10 - 11 μm $NH_3$ feature of 2MASS J0559-1404 (heavy black line) to cloudless models (Marley et al. 2002) with $T_{eff}$ = 1200 K, log g = 5.0, $K_{zz}$ = 0 $cm^2/s$ (equilibrium; thin blue line) and $K_{zz}=10^2$ $cm^2/s$ (non-equilibrium; thin red line). Both models have been normalized to the data. A typical 1-σ error bar is shown.

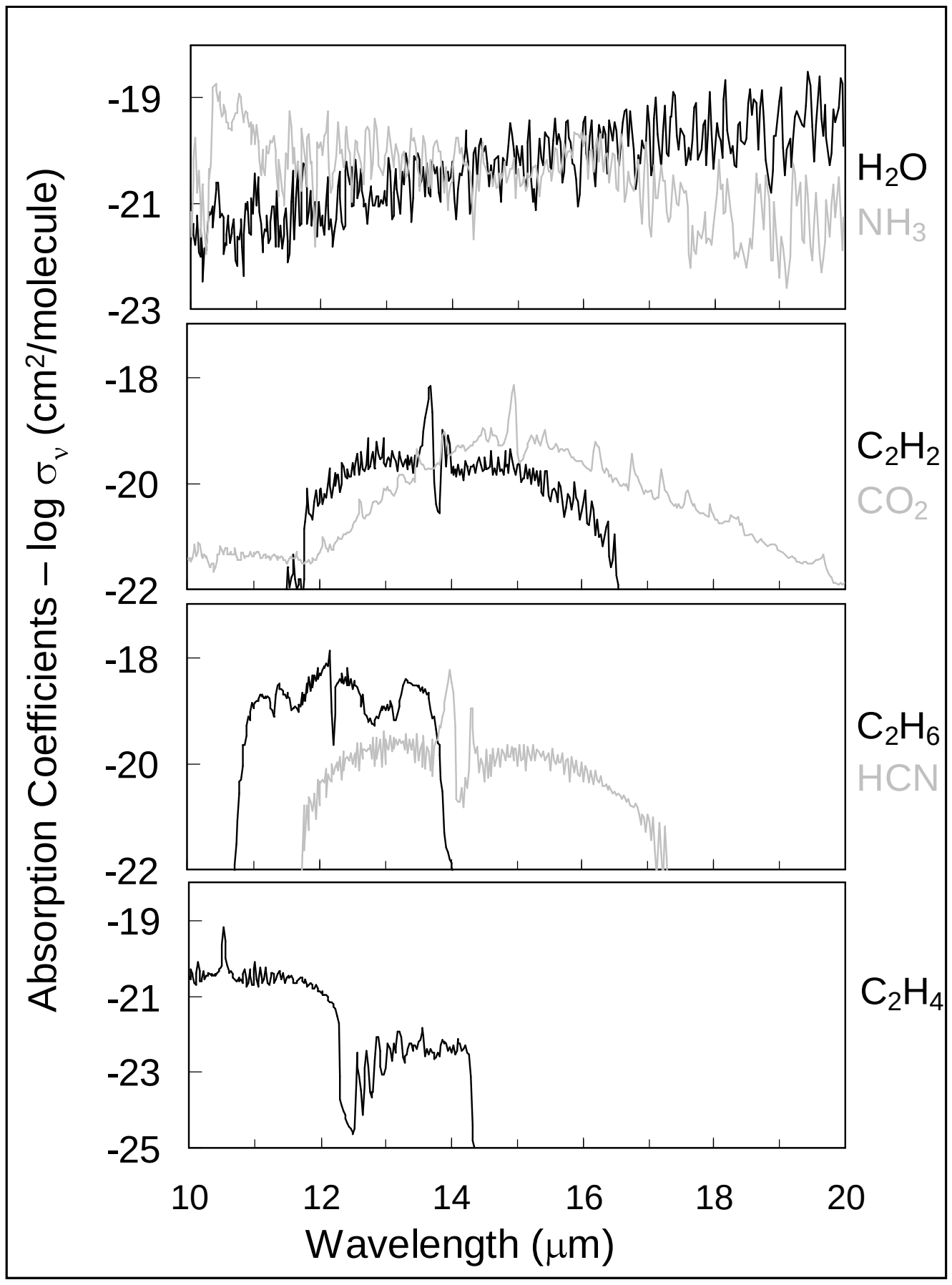

Figure 7: Absorption cross-section of various molecular species at R=600 that might be expected to occur for T=1300K and P = 1 bar.